\documentclass[12pt]{iopart} 
\usepackage[utf8]{inputenc}
\usepackage{graphicx}
\usepackage[colorlinks,citecolor=blue,urlcolor=blue,linkcolor=blue]{hyperref}
\usepackage{siunitx}

\expandafter\let\csname equation*\endcsname\relax
\expandafter\let\csname endequation*\endcsname\relax

\usepackage{amsmath}
\usepackage{iopams}
\usepackage{widetable}
\usepackage{bm}
\usepackage{cite}
\usepackage{dsfont}
\usepackage{booktabs}
\usepackage{supertabular}
\usepackage{color}
\usepackage{hyperref}
\usepackage{mathrsfs}
\usepackage{float}
\usepackage{textcomp}
\usepackage{epsfig}
\usepackage{epstopdf}
\usepackage{multirow}
\usepackage{siunitx}
\usepackage{braket}
\usepackage[titletoc]{appendix}
\usepackage{colortbl,hhline}
\usepackage{graphicx}
\usepackage{mathtools}
\usepackage{subfigure}
\usepackage{dcolumn}
\usepackage{bm}

\newcommand\newblock{\hskip .11em\@plus.33em\@minus.07em}

\begin{document}

\title[]{Photon-number parity of heralded single photons from a Bragg-reflection waveguide reconstructed loss-tolerantly via moment generating function}
\author{K.~Laiho$^{1}$,  M.~Schmidt$^{1,2}$, H.~Suchomel$^{3}$, M.~Kamp$^{3}$, S.~H\"ofling$^{3,4}$, C.~Schneider$^{3}$,  J.~Beyer$^{2}$,  G.~Weihs$^5$ and S.~Reitzenstein$^1$}
\address{$^1$Technische Universit\"at Berlin, Institut f\"ur Festk\"orperphysik, Hardenbergstr.~36, 10623 Berlin, Germany}
\address{$^2$Physikalisch-Technische Bundesanstalt, Abbestr.~2-12, 10587 Berlin, Germany}
\address{$^3$Technische Physik, Universit\"at W\"urzburg, Am Hubland,  97074 W\"urzburg, Germany}
\address{$^4$School of Physics $\&$ Astronomy, University of St Andrews, St Andrews, KY16 9SS, United~Kingdom}
\address{$^5$Institut f\"ur Experimentalphysik, Universit\"at Innsbruck, Technikerstra\ss e 25, 6020 Innsbruck, Austria}

\date{\today}

\begin{abstract}
Due to their strict photon-number correlation, the twin beams produced in parametric down-conversion (PDC) work well for heralded state generation. Often, however, this state manipulation is distorted by the optical losses in the herald and by the higher photon-number contributions inevitable in the PDC process. In order to find feasible figures of merit for characterizing the heralded states, we investigate their normalized factorial moments of the photon number that can be accessed regardless of the optical losses in the detection. We then perform a measurement of the joint photon statistics of twin beams from a semiconductor Bragg-reflection waveguide with transition-edge sensors acting as photon-number-resolving detectors. We extract the photon-number parity of heralded single photons in a loss-tolerant fashion by utilizing the moment generating function. The photon-number parity is highly practicable in quantum state characterization, since it takes into account the complete photon-number content of the target state. 
\end{abstract} 

\vspace{2pc}

\maketitle

\section{Introduction}

The process of parametric down-conversion (PDC), which generates pairs of photons, has proven to be expedient for heralding single photons \cite{Hong1986, A.I.Lvovsky2001} as well as higher photon-number states \cite{A.Ourjoumtsev2006, Waks2006, Bouillard2019}. 
When pumped with a bright laser, the twin beams from the PDC process inevitably include higher photon-number contributions, the characteristics of which have been investigated in the past via the normalized Glauber correlation functions \cite{Glauber1963, Avenhaus2010, Allevi2012}. Luckily, these correlation functions also provide valuable information of the PDC process parameters \cite{Laiho2011}. Alternatively, the higher-order moments can be extracted from the joint photon-number distribution of the twin beams \cite{Wakui2014, Harder2016}. Moreover, such direct measurements of their joint photon statistics also deliver access to the non-classicality of different classes of heralded states \cite{Sperling2017, Perina2017}.

Due to the experimental imperfections, like the optical losses  both in the state manipulation and target state detection, the measured photon statistics are often degraded and the fragile quantum features are lost \cite{Helt2017, Chesi2019}. In order to decouple the effect of these two loss contributions one can investigate the target state properties with the help of the normalized factorial moments of the photon number that can be extracted independent of the detection losses. Therefore, higher-order moments are routinely used for loss-independent verification of the quantum characteristics of radiation fields \cite{Hanbury-Brown1956, Kimble1977,  Bocquillon2009, Heindel2017}. Additionally, when combined with the loss-tolerant determination of the mean photon number of the studied target state, this method can be employed for a more direct quantum state characterization via the so-called moment generating function \cite{S.M.Barnett1997}, which provides, for example, access to the individual photon-number contributions and even to the photon-number parity \cite{Beenakker2001, Wasilewski2008, Laiho2012, Barnett2018}. 

Until today, only few (non-commercial) photo-detectors provide true photon-number resolution at the single- and few photon level. Such detectors are, for example, the visible light-photon counter \cite{Waks2006} and the superconducting transition-edge sensor (TES) \cite{Lita2008, Schmidt2018}. The latter is a very sensitive calorimeter working  at the transition edge between the superconducting phase and the normal phase.  Apart from being practicable in photon counting, that is, reconstructing the photon statistics of the measured light \cite{Y.Zhai2013, Schlottmann2017, Helversen2019}, the TES-based detectors have also been employed in spectral measurements due to their excellent energy resolution \cite{Foertsch2015}.  Besides, a quantum optical tomography of the TES has verified its ability to work as a linear photon  counter without spurious noise contributions \cite{Brida2012}.

Here, we employ a Bragg-reflection waveguide (BRW) fabricated of AlGaAs to generate broadband type-II PDC emission with a moderate photon-pair correlation \cite{Horn2012, Guenthner2015} for producing heralded single photons \cite{Belhassen2018}. Importantly, we use TES-based photon-number-resolving detectors for measuring the emitted joint photon statistics of twin beams. To easily visualize the characteristics of the heralded states we first theoretically investigate their higher-order moments in terms of the optical losses in the herald and the mean-photon number of the PDC emission \cite{DAuria2012, Rohde2015}, both of which are easily accessible in a  measurement. Thereafter, we explore the loss-tolerant reconstruction of the photon-number parity via the moment generating function providing a boundary for the reliable state reconstruction region in a real experiment.  Altogether, our investigations give insight in the suitability of the twin-beam based photon-pair sources to heralding tasks. Further, we provide a direct connection between the heralded state quality and the important parameters of the state generation and manipulation. We believe that our straightforward method for the loss-tolerant reconstruction of the photon-number parity can become practical in quantum state characterization.

\section{Theory \label{sec:theory}}
\begin{figure}[b]
\centering
\includegraphics[width = 0.45\textwidth]{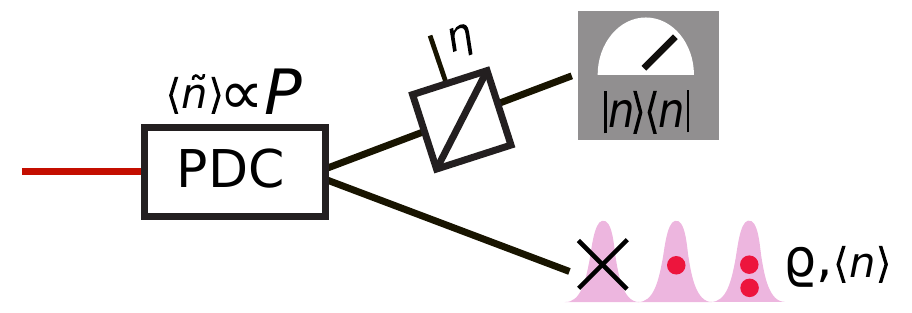}
\caption{\label{fig:scheme} Parameters involved in heralding with well-behaved PDC. The mean photon number of the PDC emission $\braket{\tilde{n}}$ is proportional to the optical pump power $P$. The twin beam, which is used as herald, is detected with a lossy photon counter that is modeled by placing a beam splitter with the transmittance $\eta$ corresponding to the heralding efficiency in front of an ideal detector that projects the state onto the photon-number basis $\ket{n}\bra{n}$. The heralded target state $\varrho$ with a mean photon number of $\braket{n}$ is prepared  in the other twin beam whenever a herald is detected.}
\end{figure}

We start by investing heralding as sketched in Fig.~\ref{fig:scheme} with a well-behaved PDC process, which generates perfectly photon-number correlated twin beams called signal and idler. This state can be expressed as
\begin{equation}
\ket{\psi} = \sum_{n =0}^{\infty} \lambda_{n}\ket{n, n}_{s \hspace{0.25ex} i},
\label{eq:pdc}
\end{equation}
in which $n$ denotes the photon number in signal ($s$) and idler ($i$) and $|\lambda|^{2}_{n}$ is the weight of each photon-number contribution. Given PDC with a mean photon number of $\braket{\tilde{n}}$, the weight distribution in Eq.~(\ref{eq:pdc})  is thermal i.e.~$|\lambda|^{2}_{n} = \braket{\tilde{n}}^{n}/(1+\braket{\tilde{n}})^{1+n}$,  if it emits into a single mode (SM) \cite{Vasilyev2000}, and turns into Poissonian i.e.~$|\lambda|^{2}_{n} = \exp{(-\braket{\tilde{n}})} \braket{\tilde{n}}^{n}/n! $, if the PDC emission is highly multimodal (MM) \cite{Haderka2005}.

In order to extract the characteristics of the heralded target state (here signal), we trace (Tr) the density matrix $\varrho_{s\hspace{0.25ex}i} = \ket{\psi}\bra{\psi}$ over the herald (here idler) as
\begin{equation}
\varrho = \frac{\textrm{Tr}_{i} \left \{ (\hat{\mathds{1}}_{s}\otimes \hat{\Pi}_{i}^{\sigma}) \ \varrho_{s\hspace{0.25ex}i}
\right\} }{\textrm{Tr}_{s\hspace{0.25ex}i} \left \{ (\hat{\mathds{1}}_{s}\otimes \hat{\Pi}_{i}^{\sigma}) \ \varrho_{s\hspace{0.25ex}i} \right \}},
\label{eq:rho}
\end{equation}
in which the operators $\hat{\Pi}_{i}^{\sigma}$ describe the photo-detector used in heralding ($\sigma$ labelling the different measurement outcomes) and form a complete set $\sum_{\sigma} \hat{\Pi}_{i}^{\sigma} = \mathds{1}$. The  photon statistics of the heralded target state is then given as a projection to the photon-number basis spanned by the photon-number states $\ket{n}$ and the probability of the $n$-th photon-number contribution can be expressed as $p(n) = \textrm{Tr}_{s} \{\varrho \ket{n}_{s \hspace{0.25ex}s \hspace{-0.25ex}}\bra{n}\}$. 

A loss-degraded photon-number-resolving detector ($\sigma=n$) used in heralding can be described by
\begin{align}
\hat{\Pi}_{i}^{n} 
= \sum_{N \ge n} \left ( \begin{array}{c}N\\n \end{array}\right) (1-\eta)^{N-n} \eta^{n}\ket{N}_{i \hspace{0.25ex}i \hspace{-0.25ex}}\bra{N},
\label{eq:phn}
\end{align}
in which $\eta$ denotes the detector efficiency. For  comparison we also examine the loss-degraded bucket detector at heralding ($\sigma =$ \emph{click} or \emph{no click}), whose properties are ideally given by
\begin{align}
\label{eq:click}
&\hat{\Pi}_{i}^{\textrm{\emph{click}}} = 1- \hat{\Pi}_{i}^{\textrm{\emph{no click}}} \quad \textrm{and} \\
&\hat{\Pi}_{i}^{\textrm{\emph{no click}}} 
= \sum_{N} (1-\eta) ^{N} \ket{N}_{i \hspace{0.25ex}i \hspace{-0.25ex}}\bra{N}. \nonumber 
\end{align}

\begin{figure}[t]
\hfill
\includegraphics[width = 0.9\textwidth]{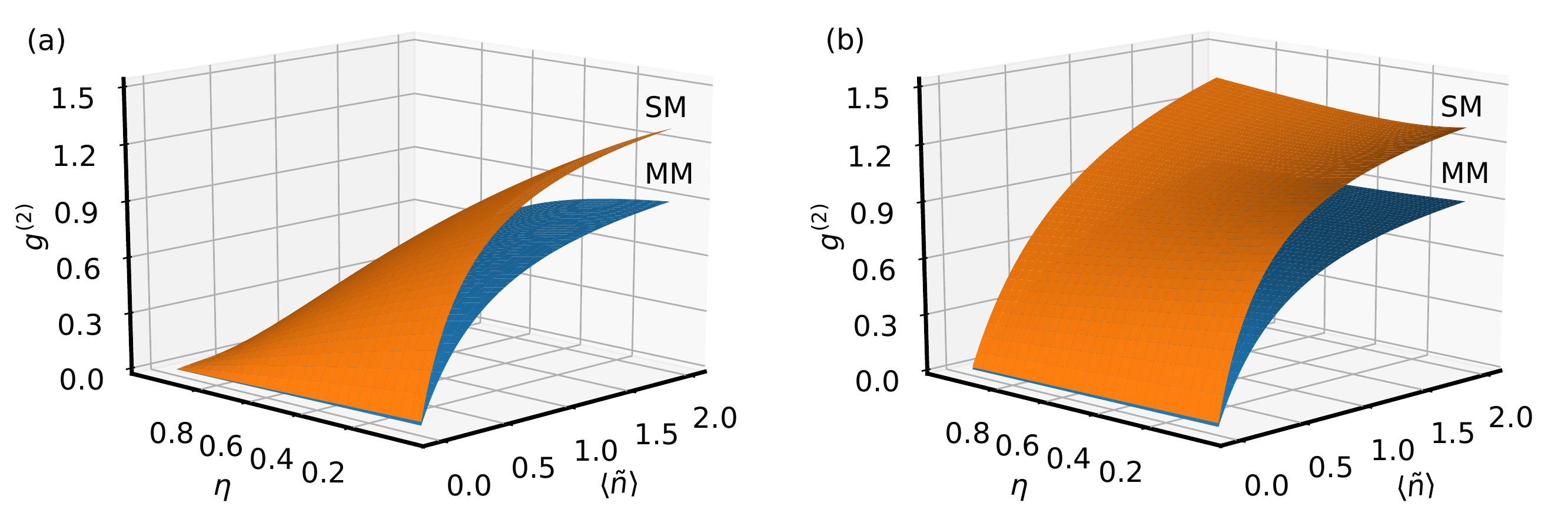}
\caption{\label{fig:one} The $g^{(2)}$-function of the heralded target state calculated via Eq.~(\ref{eq:gm}), when conditioned with a measurement of (a) one photon with a photon-number-resolving detector and (b) a single click with a bucket detector  in the herald. Since the latter can only resolve between \emph{click} and \emph{no click}, it heralds in all cases in which one or more photons impinge on it.  Orange and blue surfaces as labeled in the figure illustrate the behavior of SM and MM PDC emission, respectively.}
\end{figure}

For heralding single photons we can either choose $\sigma= 1$ [Eq.~(\ref{eq:phn})] or $\sigma = $ \emph{click} [Eq.~(\ref{eq:click})]. These are plugged into Eq.~(\ref{eq:rho}) and we calculate the resulting photon statistics. After that we employ the normalized factorial moments of the photon number for the state characterization. These moments can be extracted from the photon statistics via \cite{S.M.Barnett1997,Beenakker2001}
\begin{equation}
g^{(m)} = \frac{\braket{: \hat{n}^m :}}{\braket{\hat{n}}^{m}} =  \frac{\sum_{n} n(n-1)\dots(n-m+1) \ p(n)}{\big ( \sum_{n} n \ p(n)\big)^{m}},
\label{eq:gm}
\end{equation}
with $\hat{n}= \sum_{n} n\ket{n}\bra{n}$ being the operator delivering the mean photon number and $:\hspace{0.25ex}:$ denoting the normal-ordering of operators.
In Fig.~\ref{fig:one} we illustrate the $g^{(2)}$-function of the heralded target states in both investigated cases in terms of  the heralding efficiency and the mean photon number of the PDC emission. The differences in the target state characteristics when heralded either with a true photon-number-resolving detector in Fig.~\ref{fig:one}(a) or with a bucket detector in Fig.~\ref{fig:one}(b) are clearly visible. 
It is still possible to achieve $g^{(2)}\approx0$, with either heralding methods regardless of the heralding efficiency if the mean photon number of the PDC emission, i.e.~the pump power is kept low. 
 Additionally, the results for SM and MM PDC emission provide boundaries for the achievable $g^{(2)}$-values with well-behaved real PDC sources.  Most interestingly, however, we see that a true photon-number-resolving detector provides advantages over the bucket detector and can clearly help suppressing the effect of the higher photon-number contributions if a high heralding efficiency exceeding about 0.7 is achieved, which helps keeping $g^{(2)}\le0.5$ until $\braket{\tilde{n}}\approx 1$ for both SM and MM PDC. 
Further, we note that when regarding the generation of higher photon-number states with true photon-number-resolving detectors the situation remains very similar as for heralding single photons: thus, either a very low pump power or a high heralding efficiency is required in order to be able to generate the desired photon-number content in the heralded  target state.

The higher-order normalized factorial moments provide an interesting insight into the state characteristics, and they are connected to the so-called \emph{moment generating function} \cite{S.M.Barnett1997, Beenakker2001}
\begin{equation}
M(\mu) = \sum_{n} (1-\mu)^{n} \ p(n) = \sum_{m} \frac{g^{(m)}}{m!}(-\mu\braket{n})^{m}
\label{eq:M}
\end{equation}
given in terms of a real-valued variable $0\le\mu\le 2$ related to the operator ordering. Here, we are merely interested in reconstructing the quantum state characteristics via the right-hand side of Eq.~(\ref{eq:M}). For that purpose, apart from knowing the  higher order normalized moments also the mean photon number of the target state $\braket{n}$ needs to be measured loss-tolerantly. In general, the mean photon number can be corrected simply by dividing the measured, loss-degraded mean photon number of the target state with the efficiency, with which  it is being detected \cite{S.M.Barnett1997, Laiho2012}.
Looking further into Eq.~(\ref{eq:M}), the value of the moment generating function at $\mu = 2$ delivers the \emph{photon-number parity}, for which $-1\le M(2)\le 1$. Being the difference between the weights of the even and odd photon-number contributions, 
the photon-number parity is an important figure of merit in quantum optics \cite{Cahill1969a}. A negative photon-number parity---ideally taking the value of $M(2) =-1$ for single photons---offers a more stringent criterium of the state's non-classicality than the higher order moments do \cite{Luetkenhaus1995}. Especially, the non-classicality criterium $g^{(2)} < 1$, which can be related to the width of the photon-number distribution being sub-Poissonian, is commonly employed only to indicate to which extent photon-number contributions higher than or equal to two are existent.

\begin{figure}[t]
\hfill
\includegraphics[width = 0.82\textwidth]{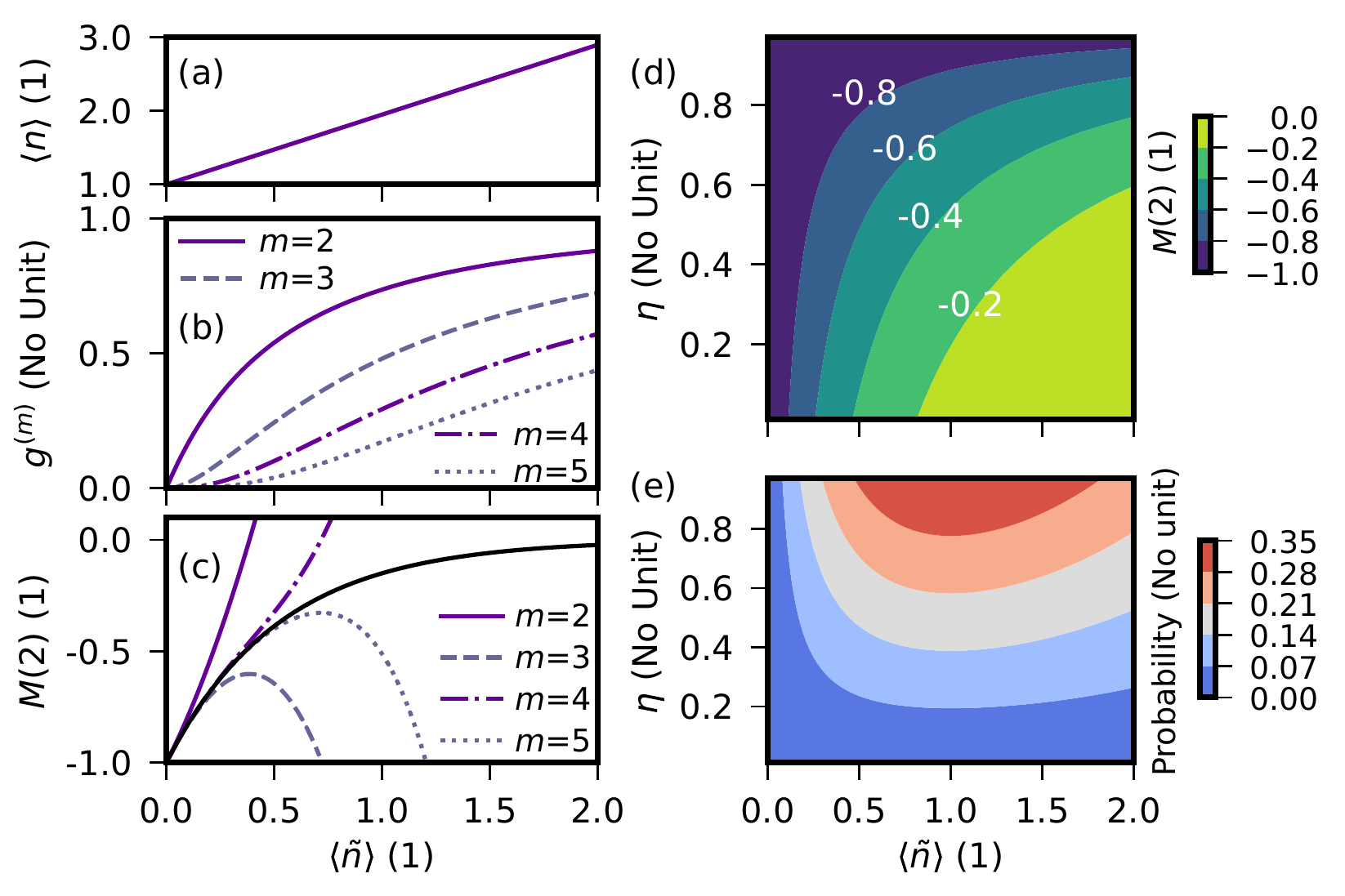}
\caption{\label{fig:two} Theoretical investigation of the photon-number characteristics of single photons heralded from MM PDC. 
(a) The mean photon number, (b) the normalized higher-order moments up to the 5th order and (c) the  photon-number parity reconstructed via the right hand side of Eq.~(\ref{eq:M}) when truncated to the $m$-th order are presented in terms of the mean photon number of the PDC emission at a heralding efficiency on the order of few percent. 
The solid black line in (c) shows the expected photon-number parity calculated via the left-hand side of Eq.~(\ref{eq:M}). 
In (d) we illustrate the photon-number parity and in (e) the target state preparation probability in terms of the heralding efficiency and  mean photon number of  PDC emission. While (d) highlights the necessity to control both these parameters in order to approach the desired value of $-1$, our results in (e) show the regions, where heralding is most efficient. }
\end{figure}

We note that extracting the photon-number properties via the alternating sum on the right side of Eq.~(\ref{eq:M}) may not necessarily converge, if the values of the normalized factorial moments grow strongly with increasing order of $m$, like for thermal states. However, when regarding certain quantum optical states such as single photons it can be practical. In Fig.~\ref{fig:two} we investigate the reconstruction of the photon-number parity of single-photon states heralded from MM PDC. We first calculate the target state characteristics at a low heralding efficiency comparable to the one in our experiment. By combining the mean photon number of the heralded state in Fig.~\ref{fig:two}(a) with the normalized factorial moments in Fig.~\ref{fig:two}(b), the photon-number parity can be reconstructed. As illustrated in Fig.~\ref{fig:two}(c) the useful reconstruction range for the photon-number parity strongly depends on the order of the highest recorded moment. Second, we perform a calculation of the photon-number parity also at higher heralding efficiencies  in Fig.~\ref{fig:two}(d), which highlights its usefulness as a figure of merit of the heralded state quality and, when  mapped in terms of the parameter space ($\eta$,$\braket{\tilde{n}}$), it denotes the regions, in which high quality single photons can be created. In contrast to the often employed photon-number fidelity, which only includes the targeted photon-number contribution, the photon-number parity  also includes information of the undesired higher photon-number contributions. Finally, Fig.~\ref{fig:two}(e) illustrates the target state preparation probability determined as $\textrm{Tr}_{si} \{ (\ket{1}_{s \hspace{0.25ex} s \hspace{-0.25ex}}\bra{1}\otimes \hat{\Pi}^{n = 1}_{i}) \  \varrho_{si}\}= \eta|\lambda_{1}|^{2}$, which denotes the probability of the successful heralding with the desired outcome in the target state. Although we found out in Fig.~\ref{fig:two}(d) that high quality target states can also be prepared at low values of $\braket{\tilde{n}}$, Fig.~\ref{fig:two}(e) shows that the state preparation in this region is very rare. 
As expected,  the region of enlarged yield lies at high heralding efficiencies close to $\braket{\tilde{n}} = 1$.

\section{Experiment}

The investigated ridge BRW has a width of  \SI{4.5}{\micro\meter}  and a length of \SI{1.7}{\milli\meter}.  Our BRW incorporates the structure first presented in~\cite{Abolghasem2009_2} and supports a collinear type-II PDC process. Thence, the pump photons in the Bragg-mode, which is a higher-order spatial mode, split  in to cross-polarized signal and idler photons both in total-internal reflection modes. Our BRW samples are grown with molecular beam epitaxy and the ridges are patterned by high-resolution electron beam lithography followed by reactive-ion plasma etching. The BRWs are etched just above the core and finally passivated with a polymer on top. The cleaved waveguide facets are kept uncoated.

\begin{figure}[t]
\hfill
\includegraphics[width = 0.9\textwidth]{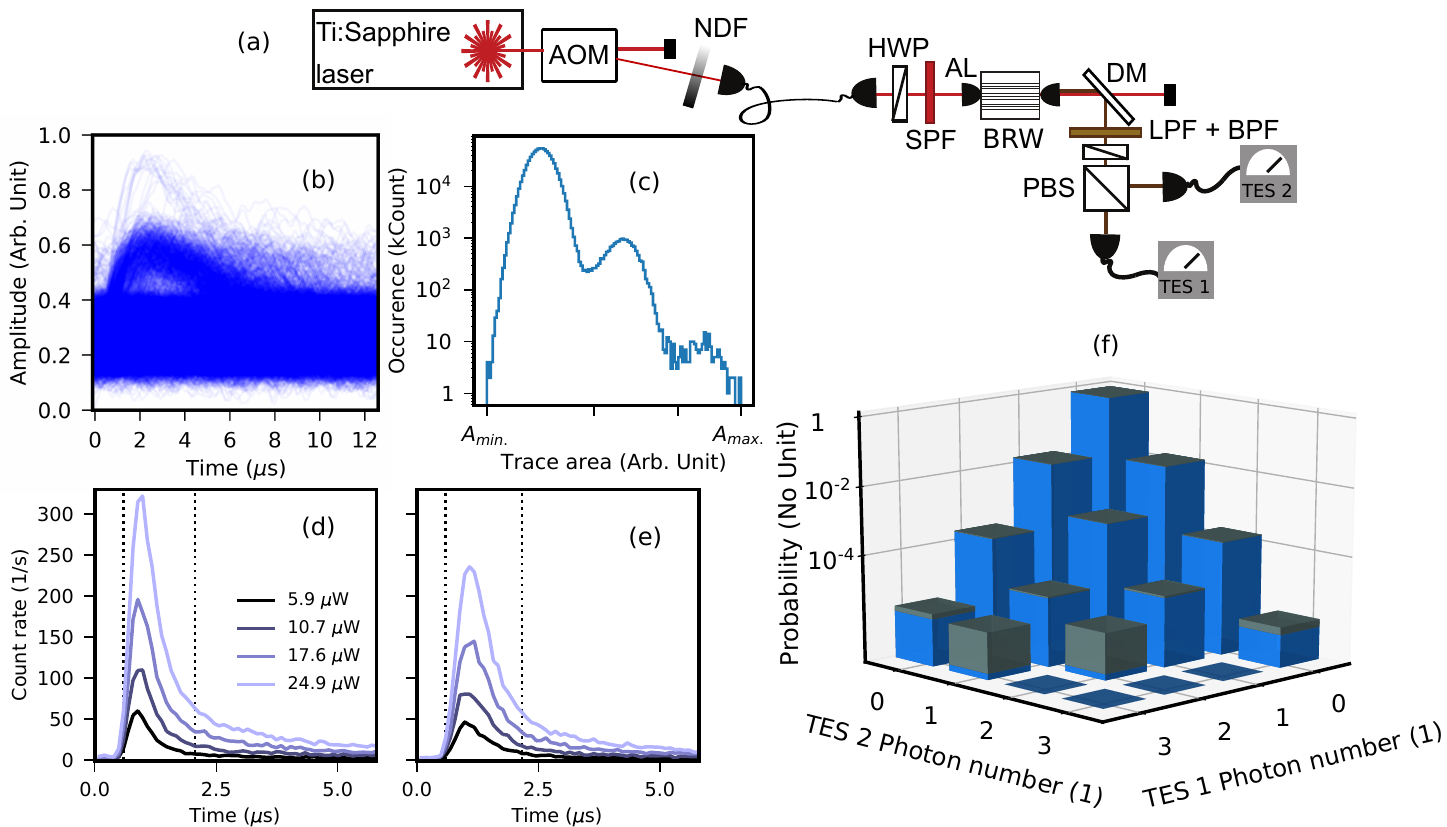}
\caption{\label{fig:setup} (a) Experimental  setup for measuring the joint photon-number distribution of twin beams from a BRW with TESs, (b) graphic presentation of  30000 raw traces  collected in about \SI{0.38}{\second} and measured at a pump power of \SI{17.6}{\micro\watt} by an individual TES, (c)  histogram of the area of the raw TES traces measured in \SI{10.4}{\second} at the pump power of \SI{17.6}{\micro\watt}, (d-e) counting rates of the two TESs with respect to time elapsed from the trigger at different pumping powers gained by post-processing the TES traces like in (b) by discriminating them with a level just above the vacuum base line and (f) the complete measured joint photon-number distribution at \SI{17.6}{\micro\watt}. The horizontal ticks in (c)  between the minimum and maximum trace area, $A_{min.}$ and $A_{max.}$ respectively, calculated from the traces including those in (b) indicate the separation between the vacuum, one- and two-photon contributions. The dashed lines in (d-e) illustrate the boundaries of the used time gates. The signal-to-noise ratio in (d-e) is on the order of 20 for both TESs being the ratio between the maximal count rate near \SI{1}{\micro\second} and  the count rate near \SI{5.5}{\micro\second}. The gray areas in (f) illustrate the measurement uncertainties in the joint photon statistics (blue bars). Abbreviations: AL = aspheric lense; AOM = acusto-optic modulator; BPF = bandpass filter; BRW = Bragg-reflection waveguide; DM = dichroic mirror; HWP = half-wave plate; LPF = long-pass filter; NDF = neutral density filter; PBS = polarizing beam spliter; SPF = short-pass filter.}
\end{figure}

In our experiment depicted in Fig.~\ref{fig:setup}(a) we use approximately \SI{1.2}{\pico\second} long Ti:Sapphire laser pulses with a central wavelength of \SI{771}{\nano\meter} and about \SI{0.5}{\nano\meter} full-width-at-half-maximum  as the pump for the PDC process. The pump beam is sent through an acusto-optic modulator (AOM) in order to reduce its repetition rate to \SI{78}{\kilo\hertz} and  the pump power is controlled with a variable neutral density filter (NDF). Thereafter, the pump beam is sent through a polarization-maintaining single-mode fiber to the BRW setup. We use a half-wave plate (HWP) and a removable sheet polarizer (not shown) for selecting the proper pump beam polarization. After passing through a short pass filter (SPF), the pump beam is coupled to the BRW for producing PDC with an aspheric lens (AL) having a focal length of approximately \SI{3}{\milli\meter}. 

Another AL collimates the PDC emission from our BRW, after which the pump beam is separated from the beam path with a dichroic mirror (DM). Thereafter,  the PDC emission is sent through a long pass filter (LPF) as well as a \SI{12}{\nano\meter} broad bandpass filter (BPF) centered close to the degeneracy wavelength of \SI{1542}{\nano\meter}, which we previously measured via the second-harmonic generation. Signal and idler then pass through another HWP and get separated in a polarizing beam splitter (PBS).  Finally, we couple the individual twin beams into single-mode fibers with ALs having  about \SI{16}{\milli\meter} fixed focus, with which we achieve $\gtrsim50\%$ coupling  efficiency near \SI{1550}{\nano\meter}.  After that signal and idler are recorded with two TESs optimized for photo-detection in the telecommunication wavelength range \cite{Lita2008}. 

The two-channel photon-number-resolving detector system includes two TESs as the light-sensitive detector elements. Each TES consists of a thin film of tungsten embedded in a multilayer structure, in which the individual photons are absorbed \cite{Lita2010}. Our TESs are kept in a demagnetization refrigerator close to \SI{100}{\milli\kelvin} with a holding time of about 25h \cite{Schmidt2018}. The individual TESs reach detection efficiencies above \SI{60}{\percent} and \SI{70}{\percent}, which were measured with a modulated continuous-wave laser in the telecommunication wavelengths. Finally,  the TESs responses are amplified with superconducting quantum interference devices \cite{Drung2007} before the detected traces are digitized with an analog-to-digital converter as shown in Fig.~\ref{fig:setup}(b). A histogram of the raw events detected by an individual TES in Fig.~\ref{fig:setup}(c) clearly shows vacuum, one- and two-photon contributions in a twin beam. Further, we utilize time gates of about \SI{1.5}{\micro\second} as illustrated in Fig.~\ref{fig:setup}(d-e) to reduce background events from the rather strong photoluminescence coming from the BRW wafer material [see Fig.~\ref{fig:setup}(b)].  The electrical trigger from the AOM is delivered for the synchronization of the measured traces.  In order to acquire the joint photon statistics as illustrated in Fig.~\ref{fig:setup}(f)  we record $(6.6-9.8)\cdot 10^{6}$ traces corresponding to effective measurement times from \SI{84}{\second} to \SI{125}{\second} and resulting up to about \SI{5}{\giga B} of raw binary data. 

\section{\label{sec:results}Results}

\begin{figure}[b]
\hfill
\includegraphics[width = 0.85\textwidth]{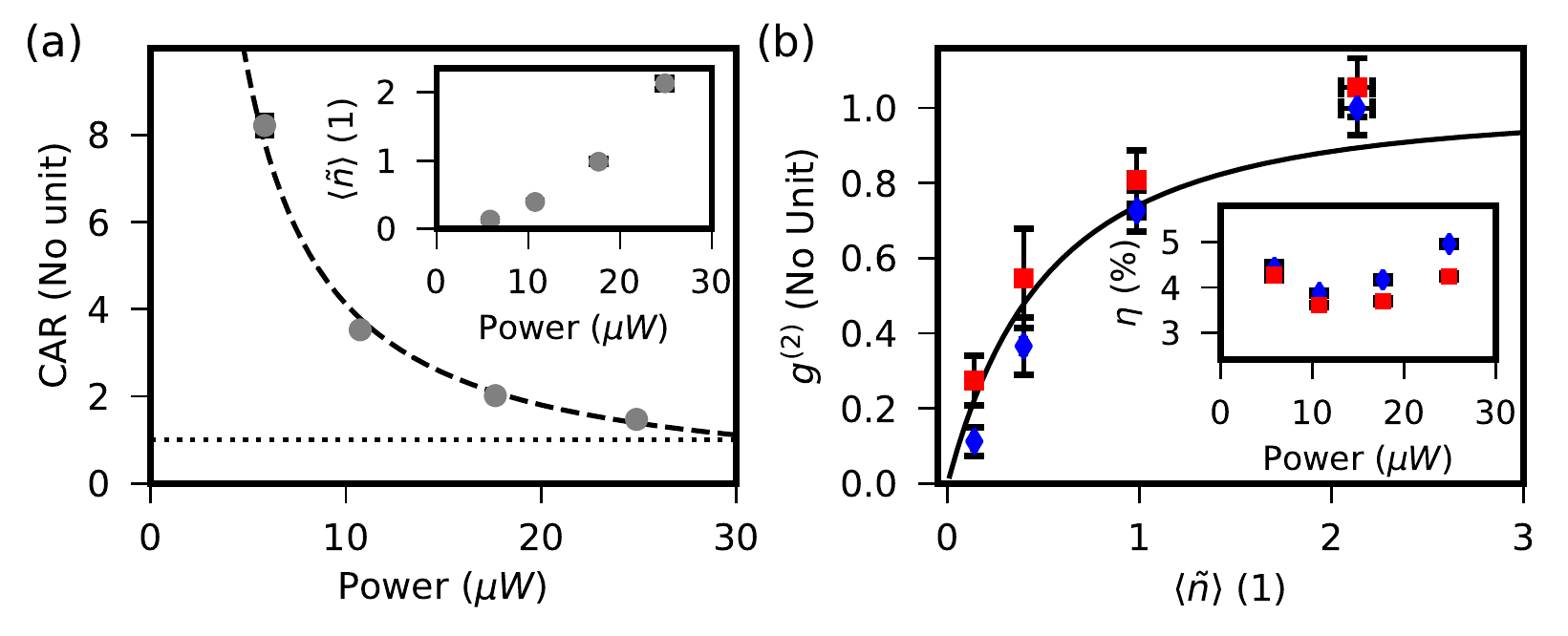}
\caption{\label{fig:CAR} (a) Measured CAR (gray circles) with respect to the power of the pump beam measured right before being coupled to the BRW. If not shown, errorbars are smaller than used symbols.  The dashed line is a fit, for which $\textrm{CAR}\propto P^{-1.19(10)}$. An exponent of $-1$ is expected for a well-behaved MM PDC process without spurious counts  \cite{Chen2018}. The dotted line highlights the value of unity expected for two completely independent photon fluxes. The inset in (a) shows the extracted mean photon number of the PDC emission vs.~pump power. (b) Evaluated $g^{(2)}$-values of the heralded states in terms of the mean photon number of the PDC emission for both signal (blue diamonds) and idler (red squares), when the other twin beam is used as herald. The solid line is the theoretical prediction for $g^{(2)}$ from Fig.~\ref{fig:two}(b). The inset in (b) shows the Klyshko efficiencies of signal and idler \cite{Klyshko1977}, which are evaluated as the ratio of coincidence counts to singles counts in the other twin beam, and are used as an estimate for the heralding efficiency.}
\end{figure}

Due to the strong spectral correlation between signal and idler our BRW produces by definition MM PDC emission \cite{Guenthner2015, Laiho2016}.  As we are only interested in the photon-number degree of freedom, our BRW  is suitable for the investigations, and we measure the joint-photon statistics of the PDC emission by varying the pump power.  To prove that photon-pair generation takes place, we first investigate the signal-idler correlation with the coincidences-to-accidentals ratio (CAR), which can be  extracted loss-independently from the measured loss-degraded joint photon statistics (see \ref{app:1}). 
Our results in Fig.~\ref{fig:CAR}(a) show, as expected, a decreasing photon-pair correlation with increasing pump power. Indeed, a value $\textrm{CAR}> 2$ is required to verify that the correlation between signal and idler is stronger than that of a thermal light source. We achieve a maximum of $\textrm{CAR} = 8.2(3)$ that is well above this limit. Moreover, the photon-pair correlation provides  a direct loss-independent access to the mean photon number of the PDC emission, which can be  estimated in the MM  case via $\textrm{CAR}=  1+1/\braket{\tilde{n}}$ \cite{Christ2011,Guenthner2015}. This is shown for our data in the inset in Fig.~\ref{fig:CAR}(a) delivering $\braket{\tilde{n}}= 0.139(5)\dots 2.13(9)$ for the range of the measured pump powers and providing in our case the lower limit for the mean photon number of the PDC emission.

We then investigate the characteristics of the higher photon-number contributions of the heralded target states gained from the joint photon statistics when conditioned on measuring one photon with the TES in  the heralding arm. In Fig.~\ref{fig:CAR}(b) we illustrate the  evaluated  $g^{(2)}$-values of the heralded states both in signal and idler in terms of the extracted mean photon number of the PDC emission. As expected, at low heralding efficiencies [see inset in Fig.~\ref{fig:CAR}(b)], the extracted values of $g^{(2)}$ increase quickly with respect to the growing mean photon number of the PDC emission. Moreover, the measured values agree well with the theoretical prediction in the MM case. In our case the preparation of high quality heralded states is strongly limited to the low values of $\braket{\tilde{n}}$, i.e.~to the low pump powers, due to the modest heralding efficiency achieved.  To improve the heralding efficiency a better suppression of spurious counts via more stringent time-gating \cite{Chen2018}, a more efficient coupling of the total-internal reflection modes into the single-mode fibers, and a reduction of the BRW's internal losses is required \cite{Pressl2015}. Nevertheless, the  preparation of single photons with close to the desired photon-number content  is still possible in our case, if the mean photon number of the PDC emission is kept low.

\begin{figure}[b]
\hfill
\includegraphics[width = 0.85\textwidth]{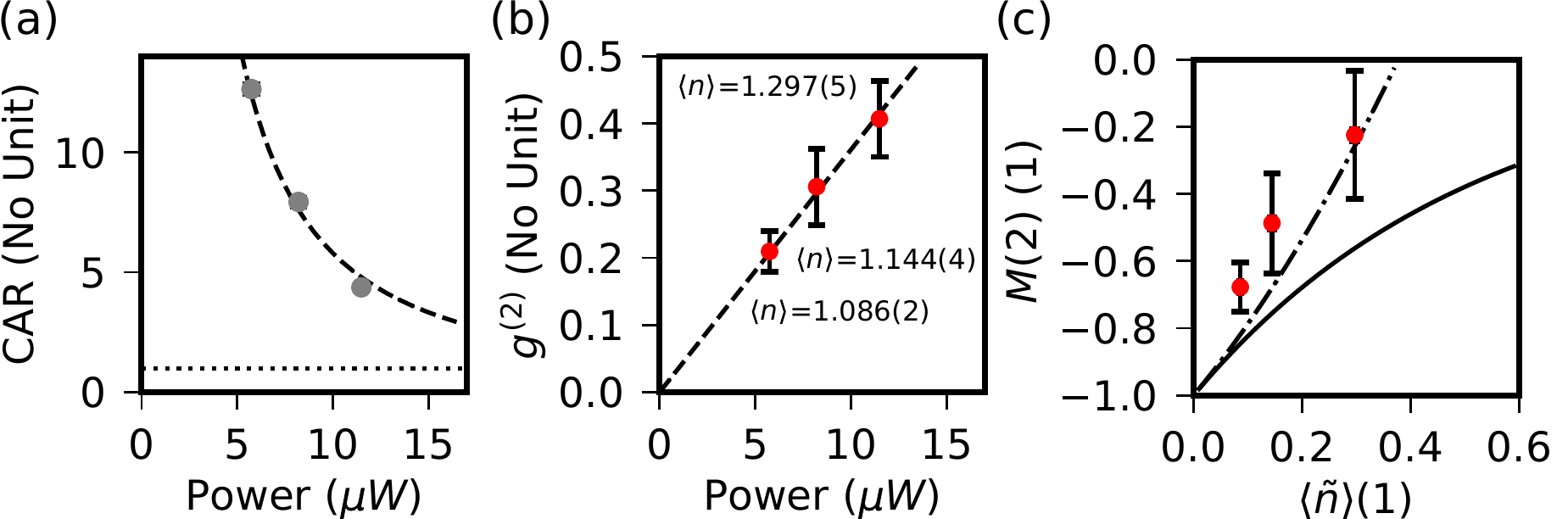}
\caption{\label{fig:parity} Experimental values (symbols) (a) for the CAR and (b) for $g^{(2)}$ of the heralded target states  with respect to pump power as well as for (c) the loss-tolerantly reconstructed photon-number parity with respect to the mean photon number of the PDC emission. The dashed line in (a) is fitted with $\textrm{CAR} \propto P^{-1.34(11)}$ and the dotted line shows $\textrm{CAR}= 1$.  The dashed line in (b) is a linear fit and the loss-corrected mean photon numbers $\braket{n}$ of the heralded target states marked are extracted from the values in (a). The reconstructed values in (c) agree well with Eq.~(\ref{eq:M}) when truncated to $m=2$ (dash-dotted line). In order to reconstruct the expected photon-number parity from Fig.~\ref{fig:two}(c) (black solid line), a measurement of normalized moments with orders much higher than two would be required. If not shown, errorbars are smaller than used symbols. }
\end{figure}

Therefore, in order to extract the photon-number parity of the heralded states via the moment generating function, we perform a second set of power dependent measurements of the joint photon statistics at low pump powers. Again, we evaluate the signal-idler correlation as shown in Fig.~\ref{fig:parity}(a). At the lowest pump power we achieve a CAR of $12.6(2)$ limiting the mean photon number of the PDC emission to $\braket{\tilde{n}}= 0.086(2)$.  In Fig.~\ref{fig:parity}(b) we illustrate the extracted $g^{(2)}$-values for the heralded target states with respect the pump power and mark their loss-corrected mean photon numbers (see \ref{app:2}). At the lowest pump power we reach the values of $g^{(2)} = 0.21(3)$ and  $\braket{n}= 1.086(2)$. Indeed, both of these figures of merit reflect the existing higher photon-number content in the heralded state.

Due to the low yield of the heralded state preparation and the limited acquisition time in our experiment, we have at the lower pump powers experimentally access only to the second-order normalized  factorial moment of the heralded states. In order to loss-tolerantly evaluate  their photon-number parity, we truncate the summation in Eq.~(\ref{eq:M}) to $m = 2$. This delivers $M(2) \approx 1-2\braket{n}+2\braket{n}^{2}g^{(2)},$ to which the data from Fig.~\ref{fig:parity}(b) is plugged and the extracted values for $M(2)$ are shown in Fig.~\ref{fig:parity}(c) with respect to the mean photon number of the PDC emission.  
Even if only the second-order normalized moment has been measured,  a negative photon-number parity can be reconstructed upto  $\braket{\tilde{n}} \lesssim 0.4$ for well-behaved MM-PDC at low heralding efficiencies as shown by the dash-dotted line in Fig.~\ref{fig:parity}(c) and our results nicely follow this tendency.
At the lowest pump power we can reconstruct the photon-number parity of $-0.68 \pm 0.08$ for the heralded state. However, one clearly sees that the  truncation of Eq.~(\ref{eq:M}), which is necessary due to the experimental limitations, causes distortion from the theoretical expectation. 

\section{Conclusions}
We measured the joint photon statistics of twin beams generated via parametric down-conversion in Bragg-reflection waveguides with transition-edge sensors, which are true photon-number-resolving detectors. We showed that the normalized factorial moments of the photon number are expedient for providing an easy access to the heralded state characteristics in terms of the  parameter space spanned by the mean photon number of the PDC emission and the heralding efficiency. By combining the measured second-order normalized moment of the heralded single photon with its mean photon number corrected for experimental imperfections, we were able to reconstruct its negative photon-number parity. We believe our experimental method can become powerful in order to loss-tolerantly investigate the characteristics of also more sophisticated states, like higher photon-number states. Additionally, our scheme is not limited to true photon-number-resolving detectors but can also be implemented with  bucket detectors connected to cascaded optical beam-splitter networks, which are also capable of measuring higher-order moments.

\section*{Acknowledgements}
We acknowledge support from the Austrian Science Fund (FWF): J-4125-N27 and I-2065-N27, the German Research Foundation (DFG): RE2974/18-1 and SCHN1376/2-1 and the State of Bavaria. The work reported in this paper was partially funded by project EMPIR 17FUN06 SIQUST. This project has received funding from the EMPIR programme cofinanced by the Participating States and from the European Union’s Horizon 2020 research and innovation program.

We gratefully thank  F.~Gericke for laboratory support,  A.~Wolf and S.~Kuhn for assistance during 
the sample growth and fabrication, M.~López from PTB Braunschweig for lending us a tunable telecom continuous-wave  laser, and A.~E.~Lita and S.~W.~Nam from NIST, USA, for providing us the transition-edge sensor chips.

\section*{References}

\appendix
\section{\label{app:1} Extracting signal-idler photon-number correlation from the joint photon statistics}

The higher-order normalized moments can be extracted directly from the measured joint photon statistics of twin beams. We calculate them directly via \cite{Avenhaus2010}
\begin{align}
g^{(m,n)} = \frac{\braket{: \hat{n}_{s}^{m}  \hat{n}_{i}^{n}  :}}{\braket{\hat{n}_{s} }^{m} \braket{ \hat{n}_{i} }^{n}},
\label{eq:g_mn}
\end{align}
in which $(m,n)$ denotes the order of the correlation and $\hat{n}_{s}$ and $\hat{n}_{i}$ are the mean photon-number operators in the  signal and idler arms, respectively.  We evaluate the moments in the photon-number basis, and therefore assume that the density matrix of the two-mode state can be written as $\varrho _{s,i}= \sum_{k, k^{\prime}} \sum_{l, l^{\prime}} c_{k, k^{\prime}, l, l^{\prime}}\ket{k}_{s \hspace{0.5ex} s}\hspace{-0.25ex} \bra{k^{\prime}}  \otimes \ket{l}_{i \hspace{0.5ex} i}\hspace{-0.25ex} \bra{l^{\prime}}$. The joint photon statistics can then be expressed as $P(k,l) = \textrm{Tr} \left \{  \varrho_{s,i} \ket{k}_{s \hspace{0.5ex} s}\hspace{-0.25ex}\bra{k} \otimes \ket{l}_{i \hspace{0.5ex} i}\hspace{-0.25ex}\bra{l} \right \} = c_{k, k, l, l} = \tilde{c}_{k,l}$. Due to the normalization $\sum_{k,l} P(k,l) = \tilde{c}_{k,l} =1$.

We can evaluate the higher-order moments  by re-writing $ \hat{n}_{s}= \hat{a}^{\dagger}\hat{a}$ and $ \hat{n}_{i}= \hat{b}^{\dagger}\hat{b}$ with $\hat{a}$ ($\hat{a}^{\dagger}$)  and $\hat{b}$ ($\hat{b}^{\dagger}$) being the photon annihilation and creation operators in signal and idler arms. The expectation values involved in Eq.~(\ref{eq:g_mn}) can the be evaluated with the help of
\begin{align}
\braket{: \hat{n}_{s}^{m}  \hat{n}_{i}^{n} : } &= \textrm{Tr} \left \{ \varrho_{s,i} \ \hat{a}^{ m}  \hat{b}^{\dagger n}  \hat{b}^{n}  \hat{a}^{m}  \right \} \nonumber \\
& = \sum_{k, k^{\prime}} \sum_{l, l^{\prime}} c_{k, k^{\prime}, l, l^{\prime}} ~_{s} \hspace{-0.25ex}  \braket{k^{\prime} |\hat{a}^{\dagger m} \hat{a}^{m}| k} \hspace{-1ex} ~_{s\hspace{0.5ex} i}\hspace{-0.25ex} \braket{l^{\prime}| \hat{b}^{\dagger n} \hat{b}^{n} |l} \hspace{-1ex} ~_{i} \nonumber\\
& = \sum_{k} \sum_{l} \tilde{c}_{k,l } \underbrace{k(k-1)\dots (k-m+1)}_{= \mathcal{G}(k)} \underbrace{l(l-1)\dots (l-n+1)}_{= \mathcal{G}(l)}\nonumber\\
& = \sum_{k} \sum_{l} \mathcal{G}(k) \tilde{c}_{k,l}\mathcal{G}(l),
\end{align}
the final form of which can be evaluated as a matrix multiplication. In Sec.~\ref{sec:results} we utilize  the signal-idler correlation $g^{(1,1)}$, which  corresponds to the well-known coincidences-to-accidentals ratio (CAR).  

\section{\label{app:2} Accessing the mean photon number of heralded state}

The signal-idler correlations can be very helpful in characterizing the properties of heralded target states and indeed,  they are useful  for estimating the mean photon number of the heralded states in a loss-tolerant manner. Because of the low detection efficiency, we can make use of coincidence counting for approximating it. Thus, the ratio of coincidence counts $C$ to single counts $S_{\xi}$ ($\xi = s,i$) for signal ($s$) and idler ($i$) provides a good approximation for the loss-degraded mean photon number of the heralded state, which is given by $\braket{n} ^{\mathrm{lossy}}_{s,i} \approx C/S_{i,s}$. The loss-inverted mean photon number can be extracted as $\braket{n}_{s,i} = \braket{n}_{s,i}^{\mathrm{lossy}}/\eta_{s,i}^{\prime}$ with $\eta_{s,i}^{\prime} = (C-A)/S_{i,s}$ being the \emph{effective} Klyshko efficiency of signal and idler, which accounts for the effect of higher photon-number contributions since the accidental counts $A$ are subtracted from the measured coincidences. Therefore, the loss-tolerantly determined mean photon number of the heralded state can be estimated with the help of the CAR denoted as $C/A$ via $\braket{n}_{s,i} \approx \frac{C}{S_{i,s}} \frac{S_{i,s}}{C-A} = (1-\frac{1}{C/A})^{-1}$.

\end{document}